\documentclass[aps,11pt,nofootinbib]{revtex4-1}
\pdfoutput=1
\usepackage{graphicx}
\usepackage[margin=1in]{geometry}
\usepackage{amsmath}
\usepackage{amssymb}
\usepackage{color}
\usepackage[utf8]{inputenc}
\usepackage{hyperref}
\usepackage[normalem]{ulem} 
\usepackage{cancel} 

\makeatletter
\renewcommand\@make@capt@title[2]{%
  \@ifx@empty\float@link{\@firstofone}{\expandafter\href\expandafter{\float@link}}%
   {\textbf{#1}}\@caption@fignum@sep#2\quad
}%
\makeatother

\begin{document}

\author{Sam Young$^1$}
\email{syoung@mpa-garching.mpg.de}
\author{Marcello Musso$^{2,1}$}
\email{mmusso@eaifr.org}

\affiliation{1) Max Planck Institute for Astrophysics, Karl-Schwarzschild-Strasse 1, 85748 Garching bei Muenchen, Germany,} 

\affiliation{2) ICTP-East African Institute for Fundamental Research, CST Nyarugenge Campus, University of Rwanda, Kigali, Rwanda }

\date{\today}

\title{Application of peaks theory to the abundance of primordial black holes}

\begin{abstract}
We consider the application of peaks theory to the calculation of the number density of peaks relevant for primordial black hole (PBH) formation. For PBHs, the final mass is related to the amplitude and scale of the perturbation from which it forms, where the scale is defined as the scale at which the compaction function peaks. We therefore extend peaks theory to calculate not only the abundance of peaks of a given amplitude, but peaks of a given amplitude and scale. A simple fitting formula is given in the high-peak limit relevant for PBH formation. We also adapt the calculation to use a Gaussian smoothing function, ensuring convergence regardless of the choice of power spectrum.
\end{abstract}
\maketitle


\section{Introduction}

Primordial black holes (PBHs) are black holes which may have formed in the early universe. Whilst no observations have been confirmed, there are several hints towards their existence \cite{Clesse:2017bsw}. They represent a viable dark matter candidate, and are a unique probe to constraint the small scale early universe. There are several different formation mechanisms, but we will focus here on PBHs formed from the collapse of large density perturbations. Shortly after the end of inflation, during the radiation dominated epoch of the universe, the cosmological horizon grows and perturbations which were super-horizon cross the horizon and can collapse to form a PBH.

In order to form a PBH, a density perturbation at horizon crossing must have an amplitude above some threshold, $\delta_c$, for gravity forces to overcome pressure forces and collapse. The density contrast $\delta(t,\mathbf{x})$ is the relative over-density, stated in the comoving synchronous gauge,
\begin{equation}
\delta(t,\mathbf{x}) = \frac{\rho(t,\mathbf{x})-\bar{\rho}(t)}{\bar{\rho}(t)},
\end{equation}
where $\rho$ is the energy density, and the bar denotes the background value for a flat universe, $\bar{\rho}(t)=3H^2(t)/(8\pi)$. $H$ is the Hubble parameter, and we are using natural units $c=G=1$.

The abundance of PBHs is then typically found by calculating the abundance of perturbations above this threshold value (see \cite{Green:2004wb,Shandera:2012ke,Nakama:2013ica,Young:2014ana,Carr:2017jsz,Germani:2018jgr,Yoo:2018esr,Young:2019yug,Kalaja:2019uju} amongst others). The simplest way of determining the abundance of such perturbations is by using a Press-Schechter-like) formalism, which has a simple condition that the density must be above the threshold. Peaks theory \cite{Bardeen:1985tr} introduces a further constraint, stating that compact objects (such as galaxies, or in our case, PBHs) form at peaks of the density. Applying this peak constraint to the condition for a perturbation to form a PBH gives us $\delta_\mathrm{D}^{(3)}(\vec{\nabla}\delta)\theta_H(-\nabla^2\delta)\theta_H(\delta-\delta_c)$. 

In the case of large-structure, it is also commonly assumed that compact objects also form at positions where the density is above a certain threshold,  and the mass of the object can be determined purely by its scale - defined as the largest smoothing scale at which the perturbation is above the threshold value. However, this is not the case for PBH formation, because a larger amplitude perturbation will pull in more of the surrounding material as a PBH forms, resulting in a larger PBH mass \cite{Niemeyer:1997mt,Musco:2008hv,Musco:2012au,Young:2019yug}. The mass of the PBH depends on both the amplitude and scale of the perturbation from which it formed (discussed further in section \ref{sec:mass}). 

To calculate the abundance of perturbations, we will therefore introduce another constraint requiring that perturbations have a specific scale. Previous calculations (i.e. \cite{Green:2004wb,Young:2014ana,Byrnes:2018clq,Young:2019yug,DeLuca:2019qsy}) utilising peaks theory have made the assumption that, when smoothed on a scale $R$, all density perturbations are exactly of this scale. 

In this paper, we will extend the peaks theory approach used in previous papers to account for the fact that both scale and amplitude must be accounted for. For simplicity, we will assume Gaussian statistics, although, as has been pointed out in several recent papers related to PBH abundance \cite{Young:2019yug,DeLuca:2019qsy,Yoo:2018esr}, the density contrast $\delta$ will not be Gaussian even if the curvature perturbation $\zeta$ is. However, it is expected that large peaks in the linear, Gaussian density field can be identified with peaks in the non-linear, non-Gaussian density \cite{DeLuca:2019qsy} - and so this could easily be accounted for following the methods of \cite{Young:2019yug}. Primordial non-Gaussianity has also been shown to have a significant effect on PBH abundance \cite{Shandera:2012ke,Byrnes:2012yx,Young:2013oia,Young:2015cyn}, which we will not consider here. 

We will begin by considering how the mass of a PBH depends on the perturbation from which it forms, before deriving an expression for the number density of peaks. Finally, we will describe how this can be used to determine the abundance and mass function of PBHs.

\section{Primordial black hole mass}
\label{sec:mass}
We will describe how the mass of a PBH may be determined from the initial perturbation from which it forms. Firstly, it is necessary to define the compaction function,
\begin{equation}
C(\mathbf{x},R) \equiv \frac{H^2(t)}{\pi^{3/2}R}\int\mathrm{d}^3\mathbf{y} \exp\left( -\frac{(\mathbf{x}-\mathbf{y})^2}{R^2} \right) \delta(t,\mathbf{y}) = \delta_R(\mathbf{x}),
\label{eqn:compaction}
\end{equation}

Throughout this paper, we will consider all perturbations in the super-horizon regime \cite{Young:2019osy}, where the time dependence of $H$ cancels exactly with the time dependence of $\delta$. The compaction function $C$ can be considered as equivalent to the volume-averaged time-independent component of the density contrast - which is typically referred to as $\delta_R$ in the literature. This allows the threshold for collapse to be stated in terms of the compaction, $C_c$. Note that this is different from the usual definition of the compaction function, equivalent to using a Gaussian smoothing window rather than a top-hat one (see \cite{Young:2019osy} for more discussion of the use of this function, and why there is a factor of 2 missing from the exponential in usual definition of a Gaussian function). This is to avoid divergences in the calculation, which would later appear (in the calculation of the variances).

For a perturbation centered at $\mathbf{x}$, the compaction function peaks at some scale $R$, and the perturbation length $r_m$ is defined as 
\begin{equation}
  C'\left( r_m \right)=0,  
\end{equation}
where the prime denotes a derivative with respect to $R$. For the same perturbation, the Gaussian window function returns slightly smaller values for $r_m$ and $C$ than the standard top-hat. The mass of the PBH is then given by the critical-scaling relationship \cite{Niemeyer:1997mt,Musco:2008hv,Musco:2012au,Young:2019yug},
\begin{equation}
M_{PBH} = \mathcal{K}M_H(r_m) \left( C - C_c \right)^{\gamma}.
\label{eqn:PBHmass}
\end{equation}
This follows the same form as the standard result utilising a top-hat window function, with different values for the constants: $\mathcal{K}\approx 10$, $C_c \approx 0.25$ (instead of 4 and 0.55 respectively \cite{Young:2019yug}), and $\gamma\approx 0.36$ (the details of this are given in appendix \ref{sec:PBHmass}). The horizon mass $M_H$ depends on $r_m$ as
\begin{equation}
M_H(r_m) = \left(\frac{r_m}{r_{\rm eq}}\right)^2M_{\rm eq},
\end{equation}
where $r_{\rm eq}$ and $M_{\rm eq}$ are the horizon scale and mass at the time of matter-radiation equality, respectively, where we have assumed radiation domination from PBH formation until the time of equality.

\begin{figure}
 \centering
  \includegraphics[width=0.7\textwidth]{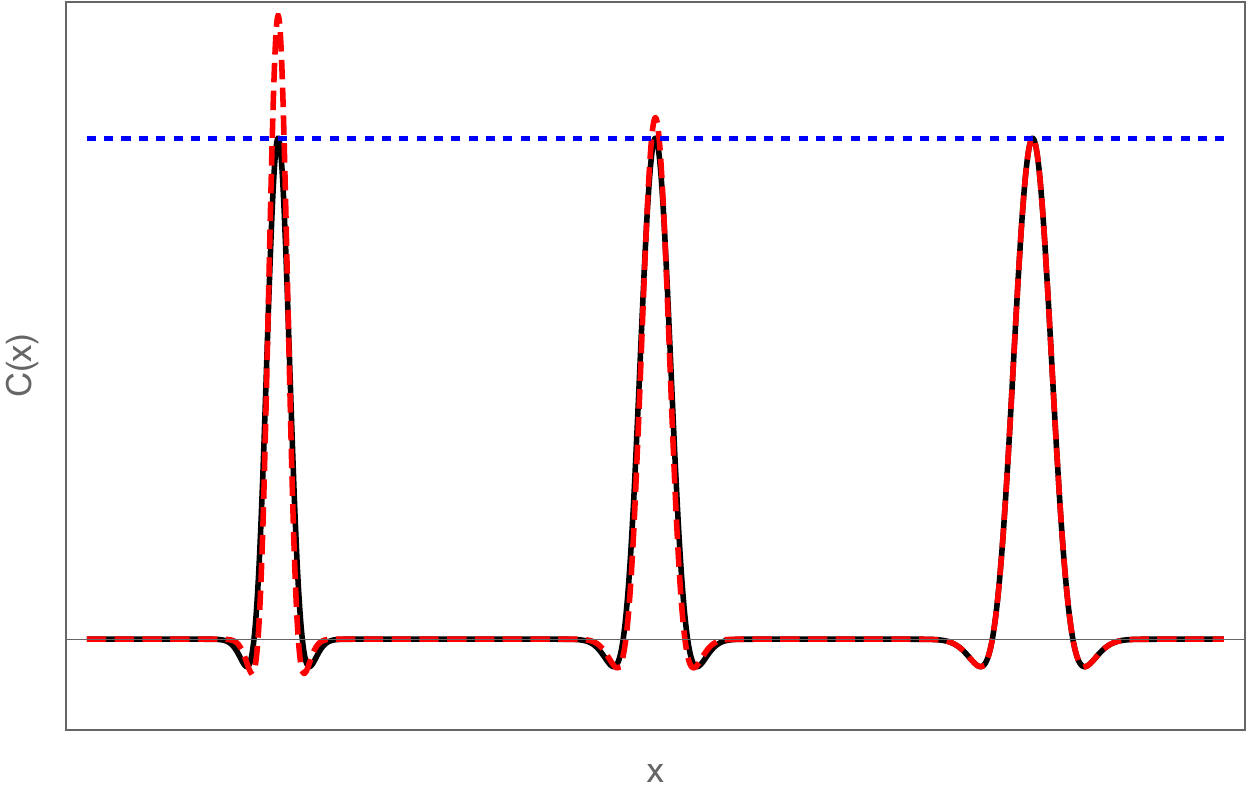} 
  \caption{ Three different peaks in the compaction of differing scales are shown, with scales $R=0.6,0.8,1$, from left-to-right in arbitrary units, and $x$ is some spatial coordinate. The black line shows the compaction when a smoothing scale $R=1$ is used, whilst the dashed red line shows the amplitude of each when smoothed on a scale corresponding to the width of that perturbation. The dotted blue line shows the collapse threshold. }
  \label{fig:threePeaks}
\end{figure}

Figure \ref{fig:threePeaks} shows a schematic plot of 3 different perturbations of different widths (given by $R=0.6,0.8,1$ in arbitrary units from left to right) and heights. If applying the excursion set, one would consider the three peaks smoothed at the largest scale at which the amplitude is above the collapse threshold. In this case, one would consider the black line - which shows the perturbations smoothed on a scale $R=1$, at which scale all 3 perturbations have exactly the critical amplitude (the amplitudes have been chosen to ensure this). The conclusion therefore, would be that all 3 form a black hole with the same mass. 

However, this would be incorrect. The dashed red line shows, instead, the amplitude of the perturbation at the smoothing scale for which the compaction is the largest. Applying then, the formula to calculate the PBH mass, equation \eqref{eqn:PBHmass}, gives the masses for the three PBHs (from left to right) $M_1 \sim 1.7 M_{\mathrm H}$, $M_2 \sim 1.6 M_{\mathrm H}$ and $M_3 \rightarrow 0$ (where $M_\mathrm{H}$ is here taken as the horizon mass when the Hubble scale $R_{H}=1$). Despite having the smallest scale, the left-hand perturbation actually forms the largest black hole, due to having the largest amplitude. The right-hand perturbation forms a vanishingly small PBH, as it has exactly the threshold density, and follows the self-similar, critical collapse described in \cite{Musco:2012au}. Care, therefore, must be taken that when PBH abundance is calculated, the number density of peaks of different scales must be evaluated at the correct smoothing scale.

\section{The peak constraint and the number density of peaks}

\subsection{Variables}
Let us now turn our attention to deriving an expression for the number density of peaks of a given height and scale. In order to derive this expression, we will introduce a large number of variables, although the final expression is much simpler. The Fourier transform of $C$, denoted by a hat, is
\begin{equation}
\widehat{C}(\mathbf{k},R) = \exp\left( -\frac{k^2R^2}{4} \right)\left( R^2H^2 \delta(t,\mathbf{k}) \right) = \widehat{\delta}_R(\mathbf{k}),
\label{eqn:Cfourier}
\end{equation}
where the second equality is included because this is the same expression as for the smoothed density-contrast (multiplied by $R^2H^2$ which cancels out the time-dependence). The Fourier transforms of the relevant derivatives of $C$ are
\begin{equation*}
\widehat{\nabla_i C}(\mathbf{k},R) =-i k_i \widehat{\delta}_R(\mathbf{k})
\;, \qquad
\widehat{\nabla^2 C}(\mathbf{k},R) = -k^2 \widehat{\delta}_R(\mathbf{k}),
\end{equation*}
\begin{equation*}
\widehat{C'}(\mathbf{k},R) = \bigg(\frac{2}{R}-\frac{k^2R}{2}\bigg) \widehat{\delta}_R(\mathbf{k}) = \frac{2}{R}\widehat{C}(\mathbf{k},R)+\frac{R}{2}\widehat{\nabla^2 C}(\mathbf{k},R),
\end{equation*}
\begin{equation}
\widehat{C''}(\mathbf{k},R) = \frac{1}{4R^2}\left( 8-10k^2R^2+k^4R^4 \right) \widehat{\delta}_R(\mathbf{k}).
\label{eqn:vars}
\end{equation}
We can see that $C'$ is a linear combination of $C$ and $\nabla^2 C$ (and $C''$ of $C$, $\nabla^2 C$ and $\nabla^2\nabla^2 C$). Each of these has the same form, a function of $k$ and $R$, multiplied by $\widehat{\delta}_R(\mathbf{k})$: $\widehat{C}_A(\mathbf{k},R) = A(\mathbf{k},r)\widehat{\delta}_R(\mathbf{k})$. The correlators of these variables can then be determined by integrating over the dimensionless power spectrum of $\delta_R$, $\mathcal{P}_{\delta_R}$
\begin{equation}
\langle C_A C_B \rangle  = \int \frac{\mathrm{d}k}{k}A(k,R)B(k,R)\mathcal{P}_{\delta_R}(k),
\label{eqn:correlators}
\end{equation}
and we will use the notation $\langle C_A C_A \rangle = \sigma_A^2$. 
We will also introduce the cross-correlation coefficient
\begin{equation}
  \gamma_{AB}\equiv 
  \frac{\langle C_A C_B \rangle}{\sigma_A\sigma_B}\,.
\label{eqn:coeffs}
\end{equation}

Because of rotational invariance, one then gets
\begin{equation}
  \langle C^2 \rangle = \sigma_0^2 \;,\quad
  \langle \nabla_iC\nabla_jC \rangle = \frac{\delta_{ij}}{3}\sigma_1^2 \;,\quad
  \langle \nabla_i\nabla_jC\nabla_k\nabla_lC \rangle = \frac{\delta_{ij}\delta_{kl}+\delta_{ik}\delta_{jl}+\delta_{il}\delta_{jk}}{15}\sigma_2^2 \;,\quad
\end{equation}
where
\begin{equation}
\sigma_n^2  = \int \frac{\mathrm{d}k}{k}k^{2n}\mathcal{P}_{\delta_R}(k)\,;
\end{equation}
all cross-correlators with an odd number of gradients vanish, while $\langle A \nabla_i\nabla_jC \rangle=- \langle \nabla_iA\nabla_jC \rangle$, where $A$ is any of the scalars $C$, $C'$, $C''$ or $\nabla^2C$. These relations imply that
\begin{equation}
  \sigma_R^2 \equiv \langle (C')^2 \rangle = 
  \frac{4}{R^2}\sigma_0^2 - 2\sigma_2^2 + \frac{R^2}{4}\sigma_2^2\,,
\end{equation}
and similar expressions hold for $\sigma_{RR}^2\equiv \langle (C'')^2 \rangle$ and $\sigma_{1R}^2\equiv \langle \vec\nabla C'\cdot \vec\nabla C'\rangle$.

As we will follow the derivation used in \cite{Lazeyras:2015giz}, we will follow the notation used there as closely as possible for ease of reference. The variables are therefore expressed in terms of variables normalised to have unit variance:
\begin{equation}
  \nu \equiv \frac{C}{\sigma_0}\,,\quad
  \eta_i \equiv \frac{\nabla_i C}{\sigma_1}\,,\quad
  \zeta_{ij} \equiv \frac{\nabla_i\nabla_j C}{\sigma_2}\,,\quad
  \eta_0 \equiv \frac{C'}{\sigma_R}\,,\quad
  \zeta_{00} \equiv -\frac{C''}{\sigma_{RR}}\,,\quad
  \zeta_{0i} \equiv \frac{\nabla_i C'}{\sigma_{1R}}\,.
\end{equation}

In the high-peak limit relevant for PBH formation, we will make the assumption $\zeta_{0i} \ll 1$, justified by the fact that, for large peaks, the physical location of a peak in $C$ is not expected to move significantly under a very small change in the smoothing scale, and therefore neglect this term when it appears.
Whilst $\nu$ and $\zeta_{00}$ are already rotationally invariant, we can define additional rotationally invariant quantities
\begin{equation}
  \eta^2 = \sum_i\eta_i^2\,,\quad
  J_1 = -\mathrm{tr}(\zeta_{ij})\,,\quad
  J_2 = \frac{3}{2}\mathrm{tr}(\bar{\zeta}_{ij}^2)\,,\quad
  J_3 = \frac{9}{2}\mathrm{tr}(\bar{\zeta}_{ij}^3)\,,
\end{equation}
where $\bar{\zeta}_{ij} \equiv \zeta_{ij} - \delta_{ij}J_1/3$.
The independent, rotationally-invariant quantities will be collectively referred to as
\begin{equation}
\mathbf{w} = \{\nu, J_1,  \zeta_{00},3\eta^2,5J_2,J_3 \},
\end{equation}
which is the same list of variables which appeared in \cite{Lazeyras:2015giz}, with one extra scalar variable, $\zeta_{00}$.

From equation \eqref{eqn:vars}, we can see that for a Gaussian smoothing window $\eta_0$ is not an independent variable but a linear combination of $\nu$ and $J_1$,
\begin{equation}
\eta_0 = \frac{1}{\sigma_R}\left( \frac{2}{R}\nu\sigma_0 -\frac{R}{2}J_1\sigma_2 \right).
\label{eqn:vR}
\end{equation}

\subsection{The peak constraint and number density of peaks}
\label{sec:npk}
Accounting for the fact that we wish to determine peaks of a given scale, the peak constraint becomes
\begin{equation}
n_{\rm pk} = \frac{3^{3/2}\sigma_{RR}}{R_*^3 \sigma_R}
\left| \begin{vmatrix}
\zeta_{00} & \frac{\sigma_{1R}}{\sigma_{RR}}\zeta_{0j} \\
\frac{\sigma_{1R}}{\sigma_{2}}\zeta_{0i} & \zeta_{ij}
\end{vmatrix}\right| 
\delta_\mathrm{D}(\eta_0) \theta_{\mathrm H}(\zeta_{00}) \delta_\mathrm{D}^{(3)}(\eta_i) \theta_{\mathrm H}(\lambda_3)\delta_\mathrm{D}(\nu-\bar{\nu}),
\end{equation}
where $R_* = \sqrt{3}\sigma_1/\sigma_2$, $\lambda_3$ is the smallest eigenvalue of $\zeta_{ij}$, $\delta_\mathrm{D}^{(n)}$ is the $n$-dimensional Dirac-delta function, $\theta_{\mathrm H}$ is the Heaviside step function, and $||\dots||$ stands for the absolute value of the determinant of the Hessian matrix. This is the Jacobian determinant of the coordinate transformation from $C'$ and $\nabla\delta$ (the Gaussian variables of the constraint) to the physical coordinates $R$ and $\mathbf{x}$.
This function returns the number of peaks in an infinitesimal volume $\mathrm{d}^3x$ with scale between $R$ and $R+\mathrm{d}R$ and height between $\nu$ and $\nu+\mathrm{d}\nu$, divided by $\mathrm{d}R\mathrm{d}\nu\mathrm{d}^3x$.
The determinant in the above equation can be factorized as
\begin{equation}
  ||\dots|| = \bigg|\zeta_{00}-\frac{\sigma_{1R}^2}{\sigma_{RR}\sigma_2}\zeta_{0i}\zeta_{ij}^{-1}\zeta_{0j}\bigg|
  \big|\det(\zeta_{ij})\big|\,.
\label{eqn:det}
\end{equation}

In the high peak limit we are interested in, the scalars $\zeta_{00}$ and $J_1=-\mathrm{tr}(\zeta)$ are both very large, since they correlate with $\nu$ and $\nu=\bar\nu\gg1$. Conversely, the traceless part $\bar\zeta_{ij}$ and the vector $\zeta_{0i}$ remain of order $O(1)$. Hence, the 3-D Hessian matrix $\zeta_{ij}$ can be approximated by $-(J_1/3)\delta_{ij}$, and the term $\zeta_{0i}\zeta_{ij}^{-1}\zeta_{0j}$ is negligible (it is suppressed twice) compared to $\zeta_{00}$.
Ignoring the $\phi_i$ terms, the peak constraint then becomes
\begin{equation}
n_{\rm pk} = \frac{3^{3/2}\sigma_{RR}}{R_*^3 \sigma_R}\left|\zeta_{00}\right|\left| \mathrm{det}(\zeta_{ij})\right| \delta_\mathrm{D}(\eta_0) \theta_{\mathrm H}(\zeta_{00}) \delta_\mathrm{D}^{(3)}(\eta_i) \theta_{\mathrm H}(\lambda_3)\delta_\mathrm{D}(\nu-\bar{\nu}).
\end{equation}

Technically, we should worry about the PBHs which form from such peaks being inside the radius of larger PBHs which form later - the so-called ``cloud-in-cloud'' problem. However, this has been found to have a negligible effect on the PBH abundace, due to the rarity of PBHs and the very small probability of such an occurence \cite{MoradinezhadDizgah:2019wjf,DeLuca:2020ioi}.

The average number density can be expressed by integrating over the probability density function (PDF) of $\mathbf{w}$, $P(\mathbf{w})$,
\begin{equation}
\bar{n}_{\mathrm{pk}} = \int \mathrm{d}\mathbf{w}P(\mathbf{w})n_{pk}\,.
\label{eqn:averagePeaks}
\end{equation}
Up until this point, the expression is completely general, and no assumptions have been made about Gaussianity. Assuming that the density contrast (and therefore the compaction) is Gaussian, the PDF can be expressed as \cite{Lazeyras:2015giz}
\begin{multline}
P(\mathbf{w}) \mathrm{d}\mathbf{w} = \mathcal{N}(\nu,J_1,\zeta_{00})\mathrm{d}\nu\mathrm{d}J_1\mathrm{d}\zeta_{00}\\
\times\left[\chi_3^2(3\eta^2)\mathrm{d}(3\eta^2)\chi_5^2(5J_2)\mathrm{d}(5J_2)\frac{1}{2}\theta_H\left( 1 - \frac{J_3^2}{J_2^3} \right)\mathrm{d}\left( \frac{J_3}{J_2^{3/2}}\right) P(\mathbf{\Omega})\mathrm{d}\mathbf{\Omega}\right],
\end{multline}
where $\mathcal{N}(\nu,J_1,\zeta_{00})$ is a 3-variate normal distribution, $\chi_n^2(x)$ is a $\chi^2$ distribution with $n$ degrees of freedom, $\mathbf{\Omega}$ is vector of the angular variables, with $P(\mathbf{\Omega})$ its PDF. The terms inside the square brackets in the above equation are unchanged from the standard peaks theory calculation, and the integral over the $3\eta^2,5J_2,J_3$ terms and the factor $3^{3/2}$ in equation \eqref{eqn:averagePeaks} gives the function $f(J_1)$ defined in reference \cite{Bardeen:1985tr},
\begin{multline}
f(J_1) = \sqrt{\frac{2}{5\pi}}\left[ \left( \frac{J_1^2}{2}-\frac{8}{5} \right)\exp\left( \frac{-5J_1^2}{2} \right)+\left( \frac{31J_1^2}{4}+\frac{8}{5} \right)\exp\left( \frac{-5J_1^2}{8} \right) \right] \\
 +\frac{1}{2}\left( J_1^3 - 3J_1 \right) \left[ \mathrm{Erf}\left( \sqrt{\frac{5}{2}}J_1 \right)+ \mathrm{Erf}\left( \sqrt{\frac{5}{2}}\frac{J_1}{2} \right) \right].
\label{eqn:f}
\end{multline}

Making use of equation \eqref{eqn:vR}, the Dirac-delta function $\delta_\mathrm{D}(\eta_0)$ can be rewritten as
\begin{equation}
  \delta_\mathrm{D}(\eta_0) =\frac{2\sigma_{R}}{R\sigma_2}
  \delta_\mathrm{D}\bigg(\frac{4\sigma_0}{R^2\sigma_2}\nu - J_1\bigg),
\end{equation}
and for convenience we will define $\bar{J}_1 \equiv 4\sigma_0\bar\nu/(R^2\sigma_2)$. Finally, integrating over the remaining Dirac-delta functions and Heaviside step functions yields
\begin{equation}
  \bar{n}_{\mathrm{pk}} =
  \frac{2\sigma_{RR}\sigma_2^2}{3^{3/2}(2\pi)^{3/2}\sigma_1^3 R}
  f(\bar{J}_1) \mathcal{N}(\bar{\nu},\bar{J}_1)
  \int_{0}^{\infty}\mathrm{d}\zeta_{00}\, \zeta_{00}\, \mathcal{N}(\zeta_{00}|\bar{\nu},\bar{J}_1).
\label{eqn:barN}
\end{equation}
where $\mathcal{N}(\zeta_{00}|\bar{\nu},\bar{J}_1) \equiv \mathcal{N}(\bar{\nu},\bar{J}_1,\zeta_{00})/\mathcal{N}(\bar{\nu},\bar{J}_1)$ is the conditional PDF of $\zeta_{00}$ given $\bar\nu$ and $\bar J_1$. We give its exact expression in Appendix \ref{app:PDF}. Equation \eqref{eqn:barN} can be considered as the main result of this paper. Its prefactor has dimensions of $[R^{-4}]$, consistently with the fact that we are computing a differential number density per unit of smoothing scale.

In the high-peak limit relevant for PBH formation, we can approximate $f(\bar{J}_1) \approx \bar J_1^3 = 64\sigma_0^3\bar{\nu}^3/(R^6\sigma_2^3)$. In this same limit, the conditional PDF $\mathcal{N}(\zeta_{00}|\bar{\nu},\bar{J}_1)$ becomes highly concentrated around its mean value $\langle\zeta_{00}|\bar{\nu},\bar{J}_1\rangle$, and the integral in equation \eqref{eqn:barN} tends to $\alpha\bar\nu\mathcal{N}(\bar{\nu},\bar{J}_1)$, where $\alpha\equiv\langle\zeta_{00}|\bar{\nu},\bar{J}_1\rangle/\bar\nu$ depends on $R$ and on $\mathcal{P}(k)$ but no longer on $\bar\nu$. We give the exact expression for $\alpha$ in equation \eqref{eqn:alpha} (though $\alpha$ is close to unity except for narrow power spectra).
Using the explicit expression for $\mathcal{N}(\bar{\nu},\bar{J}_1)$, which we derive in detail in Appendix \ref{sec:highPeak}, we can, to high accuracy, approximate the number density of peaks as
\begin{equation}
  \bar n_{\mathrm{hi-pk}} (\nu)= \frac{16\sqrt{2}}{3^{3/2}\pi^{5/2}}\frac{\sigma_{RR}\sigma_0^3}{\sigma_2\sigma_1^3 R^7\sqrt{1-\gamma_{0,2}^2}}\alpha \nu^4\exp\left( - \frac{1+\frac{16\sigma_0^2}{R^4\sigma_2^2}-\frac{8\sigma_0\gamma_{0,2}}{R^2\sigma_2}}{1-\gamma_{0,2}^2} \frac{\nu^2}{2}\right),
\label{eqn:hiPeak}
\end{equation}
where the bars have now been dropped for simplicity, and $\gamma_{0,2}$ is the cross-correlation coefficient of $\nu$ and $J_1$. Corrections to this result are of the same order of the terms that were already neglected from equation \eqref{eqn:det}.
 Equation \eqref{eqn:hiPeak} can be compared to the average number density of high peaks, $\tilde{n}_{\mathrm{hi-pk}}$, obtained using un-modified peaks theory:
\begin{equation}
\tilde{n}_{\mathrm{hi-pk}} = \frac{1}{3^{3/2}(2\pi)^2}\frac{\sigma_1^3}{\sigma_0^3}\nu^3\exp\left( -\frac{1}{2}\nu^2 \right),
\end{equation}
which being a standard number density has dimensions of $[R^{-3}]$. This result crucially differs from ours by one power of $\nu\gg1$ and by the constant multiplying $\nu^2$ in the exponential.

We also emphasize that our derivation relied on the assumption of Gaussian statistics only for the exact expression of $f(J_1)$, equation \eqref{eqn:f}. However, the limit $f(J_1)\simeq J_1^3$ holds independently of the distribution, simply because  $|\det(\zeta_{ij})|\simeq (J_1/3)^3$ for high peaks, and all non-scalar variables are marginalized over. Therefore, the generic number density of high peaks can be obtained from equation \eqref{eqn:hiPeak} simply by replacing $\alpha$ and $\mathcal{N}(\bar{\nu},\bar{J}_1)$ with the appropriate non-Gaussian values: the abundance of very high peaks can be easily obtained once the PDF of the scalar variables is known.

\section{Discussion}
\label{sec:discussion}

Here, we have considered PBHs forming from the collapse of large-density perturbations upon horizon entry. Such PBHs form with a mass close to the horizon mass at this time - the exact mass depends on both the amplitude and scale $R$ of the perturbation from which a PBH forms. We have therefore extended peaks theory to derive an expression for the number density of peaks of a given height $\nu$ and scale $R$.

Whilst we have here assumed Gaussian statistics for the density contrast $\delta$, it has recently been discussed in a number of papers \cite{Young:2019yug,DeLuca:2019qsy,Yoo:2018esr} that $\delta$ will not have a Gaussian distribution even if the curvature perturbation $\zeta$ does. However, the easiest way to account for this is to note that high peaks in the smoothed Gaussian field correspond to peaks in the smoothed non-Gaussian field \cite{Young:2019yug,DeLuca:2019qsy}, although the recent paper \cite{Matsubara:2020lyv} may be used to describe the statistics of peak of non-Gaussian fields. In order to calculate the abundance and mass function of PBHs, the expression derived here for the number density of peaks can be substituted into the existing calculation (for example, that given in \cite{Young:2019yug}) in place of the previous expression for $\bar{n}_{pk}$ (more details for the calculation in the Gaussian case are given in appedix \ref{app:abundance}). Whilst the full derivation is non-trivial and is left for future study, we can make a simple estimate of the impact of the new expression by comparing the expressions for the number density of peaks.

\begin{figure}
 \centering
  \includegraphics[width=0.7\textwidth]{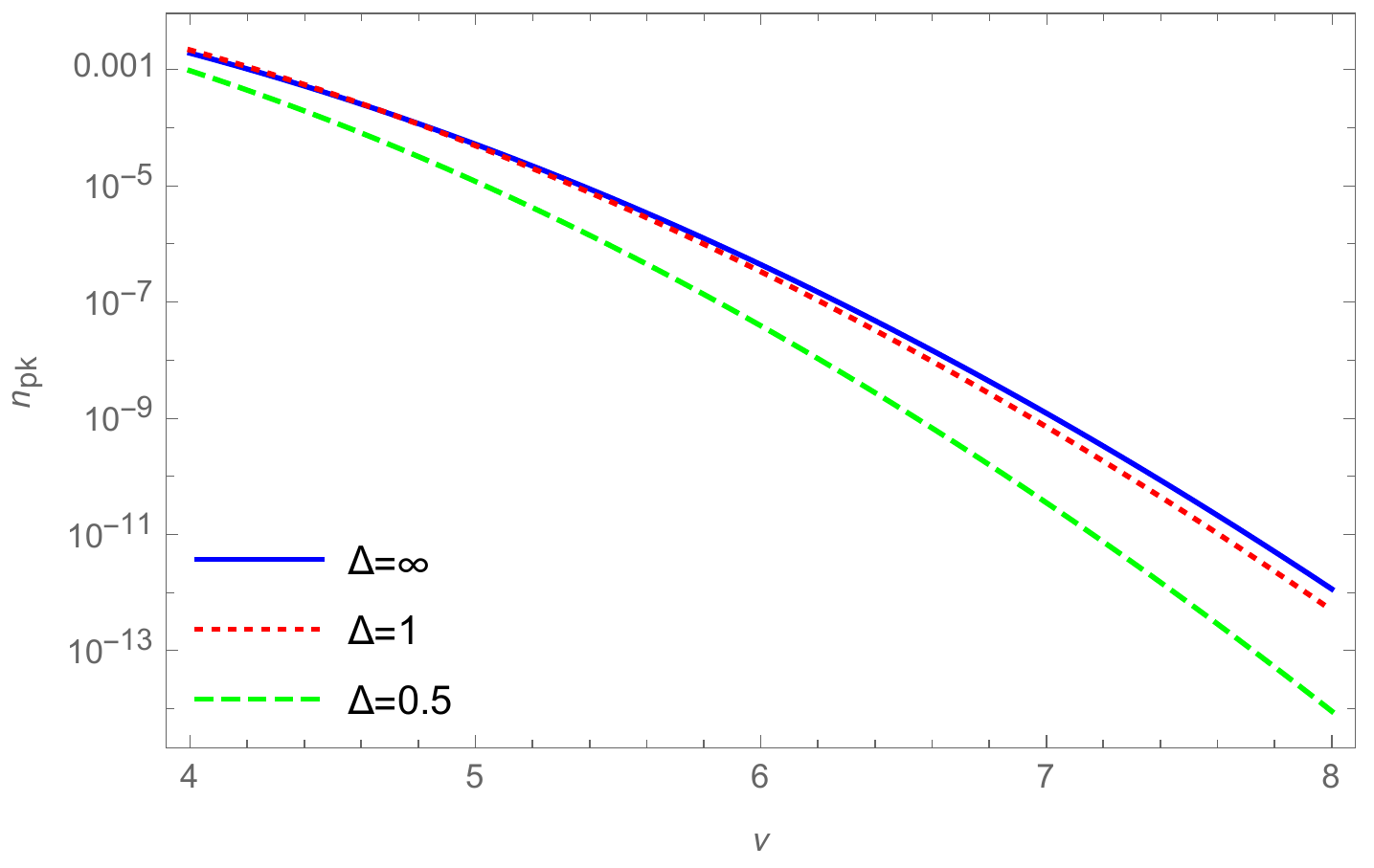} 
  \caption{ The number density of peaks is shown as a function of $\nu$ for different widths of a lognormal peak in the power spectrum. We have considered peaks with a width $R$ corresponding to the scale of the peak of the lognormal power spectrum. }
  \label{fig:width}
\end{figure}

Whether the peak number density is increased or decreased will generally depend on the form of the power spectrum and the scale of perturbations being considered. To make a comparison, we will consider a power spectrum with a lognormal peak (the form of the power spectrum is given in equation \eqref{eqn:powerSpectrumPeak}), and consider the number density of peaks on the scale corresponding to the peak of the power spectrum, $R=1/k_*$ (and we will take $R = 1/k_*=1$ for convenience). Figure \ref{fig:width} shows the number density of peaks changes as a function of $\nu$ for different widths $\Delta$ of the lognormal peak. Broader power spectra typically predict a larger number density of peaks for the same $\nu$. For the same amplitude of the power spectrum $\mathcal{A}$ we also expect the variance of perturbations $\sigma_0^2$ to be larger, so perturbations of the same amplitude $C$ will have a smaller `relative' amplitude $\nu$ associated with it (also implying a larger number of peaks (and PBHs) would form).

\begin{figure}
 \centering
  \includegraphics[width=0.7\textwidth]{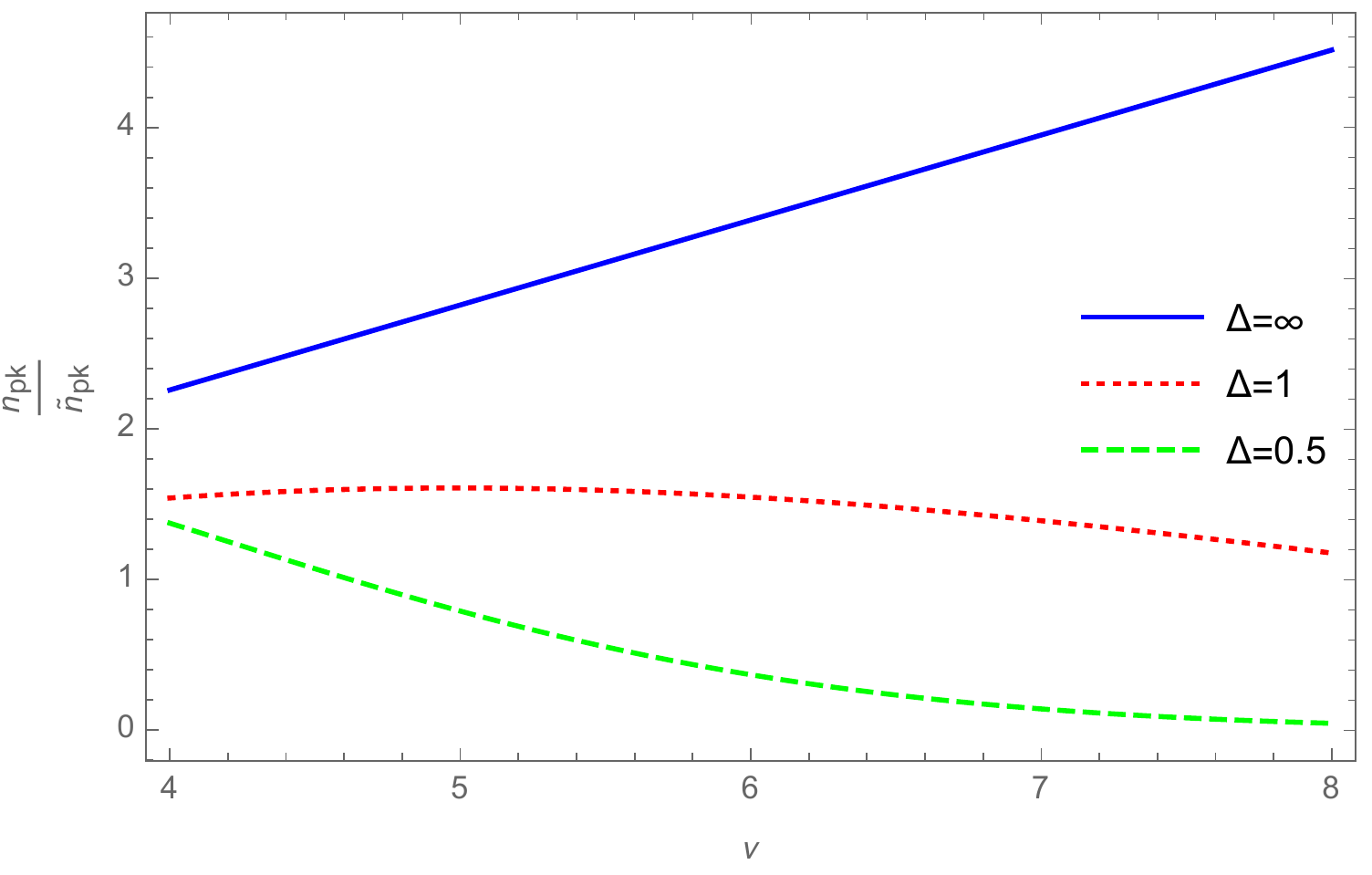} 
  \caption{ A comparison between standard peaks theory and the extended version developed in this paper is shown, for peaks in the power spectrum of different widths. Again, we are considering peaks with a width $R$ corresponding to the scale of the peak of the lognormal power spectrum. For broad peaks in the power spectrum, more peaks are predicted by a factor proportional to $\nu$. However, for narrow peaks, the new expression predicts a significantly smaller value. }
  \label{fig:method}
\end{figure}

Figure \ref{fig:method} shows a comparison between the new extended peaks theory calculation $n_{pk}$, and the previous expression $\tilde{n}_{pk}$. For broad power spectra, $\Delta\gg1$, we see an increase in the number density of peaks by a factor $\sim\nu$, whilst a significant decrease is seen for very narrow peaks, $\Delta\ll1$. For peaks in the power spectrum with $\Delta \approx 1$, approximately the same number of peaks is predicted, by a factor of order unity.

\section*{Note added}
As this paper was being written, two similar papers appeared on the arXiv, although following different methodology. Reference \cite{Suyama:2019npc} appeared first, and provides a similar argument for the inclusion of the condition that only perturbations where the compaction function peaks on a specific scale should be included when one wishes to calculate the mass function of PBHs, but does not make a comparison of the effect of including this term.

Reference \cite{Germani:2019zez} appeared shortly thereafter, and performed a fuller calculation. That paper concluded that previous (linear) calculations over-predict the abundance of PBHs for broad power spectra, seemingly in contrast with the results presented here. However, a direct comparison of these conclusions is complicated due to the different methods being used in each paper. We also note that, in this paper, we have utilised a Gaussian smoothing function, which ensures convergence of the variances regardless of the form of the power spectrum.



\section*{Acknowledgements}
SY is funded by a Humboldt Research Fellowship for Postdoctoral Researchers. Fabian Schmidt and members of the physical cosmology group the Max Planck Institute for Astrophysics are thanked for useful discussions.

\appendix

\section{Primordial black hole mass when a Gaussian smoothing function is used}
\label{sec:PBHmass}

Using the standard definition of the compaction function (i.e. a top-hat smoothing function rather than a Gaussian function, introduced in the context of PBH formation in \cite{Musco:2018rwt}), the mass of a PBH arising from a perturbation has been shown to be \cite{Niemeyer:1997mt,Musco:2008hv,Musco:2012au,Young:2019yug}
\begin{equation}
M_{PBH} \approx \mathcal{K}_{TH}M_H(r_{TH}) \left( C_{TH} - C_{c,TH} \right)^{\gamma}.
\label{eqn:tophatMass}
\end{equation}
where $M_{\mathrm H}$ is the horizon mass, $\mathcal{K}\approx 4$, $C_c\approx 0.55$ and $\gamma \approx 0.36$, although it is noted that the exact values depend on the profile shape of the perturbation. The horizon mass is the mass of the unperturbed universe contained within the Hubble radius, when the Hubble radius is equal to the length scale of the perturbation, $r_{TH}$. The subscript TH has been used to note that these are the values when a top-hat smoothing function is used, G will be used to indicate the values when a Gaussian smoothing function is used. We will now discuss how this formula should be modified if a Gaussian smoothing function is used instead.

The shape of the primordial perturbations is not known, but the ``average'' profile shape can be calculated from the statistics of the power spectrum. We will treat the large, rare perturbations which may form PBHs as spherically symmetric \cite{Bardeen:1985tr}, and therefore describe the perturbation shape using only the radius from the centre of the perturbation $r$. For a wide range of power spectrum shapes, it has been shown that the central region of large perturbations (the relevant part for PBH formation) is well approximated by a sinc function \cite{Young:2019osy}
\begin{equation}
\delta(r) = A \frac{\mathrm{sin}(r)}{r},
\end{equation}
where we are ignoring the time-dependence of the perturbation, and the radial coordinate $r$ is defined in arbitrary units (i.e. Different scale perturbations can be defined by changing the units). 

Using a top-hat or Gaussian smoothing function to calculate $r_m$ gives $r_{TH} \approx 2.74$ or $r_G = 2$ respectively. The Gaussian function gives a smaller characteristic scale for the perturbation, which then predicts a horizon mass which is smaller by a factor $(r_G/r_{TH})^2$,
\begin{equation}
M_H(r_{TH}) = \left(\frac{r_{TH}}{r_G}\right)^2 M_H(r_{G}).
\end{equation}
Likewise, the calculated amplitude of the perturbations (stated in terms of the compaction function at its maximum) will be different depending on the smoothing function used, but are proportional in the ratio $C_{TH} \approx 2.17 C_{G}$ (where the factor 2.17 is simple to find numerically).

Substituting these values into the expression for the PBH mass, equation \eqref{eqn:tophatMass} gives
\begin{equation}
M_{PBH} \approx 4 \times \left( \frac{2.74}{2} \right)^2M_H(r_{G}) \left( 2.17 C_{G} - 2.17 C_{c,G} \right)^{0.36} \approx 10 M_H(r_{G}) \left( C_{G} - 0.25 \right)^{0.36},
\end{equation}
corresponding to the values given in equation \eqref{eqn:PBHmass}.

\section{Correlation factors}

In this section, we will further discuss the correlation factors which impact the calculation of peak number density. The correlation factors are normally expressed in terms of moments of the power spectrum, and we include this here for completeness. The $n^{th}$ moment of the power spectrum is defined as
\begin{equation}
\sigma_n^2 = \int\limits_0^\infty \frac{\mathrm{d}k}{k} k^{2n} \mathcal{P}_{\delta_R},
\end{equation}
where $\delta_R$ is equivalent to the compaction function, as defined in equation \eqref{eqn:Cfourier}. The power spectrum $\mathcal{P}_{\delta_R}$ can be considered equivalent to density power spectrum smoothed on a scale $R$, and rescaled by a factor $R^4H^4$ (the rescaling can, in turn, be considered as a linear extrapolation to the amplitude at horizon entry).

The relevant correlation factors are the correlation factors of $\nu=\frac{C}{\sigma_0}$, $J_1=\frac{C_2}{\sigma_2}$ and $\zeta_{00}=\frac{C_{RR}}{\sigma_{RR}}$. Therefore, it is desirable to express $C_{RR}$ in terms of spatial derivatives of $C$, which, in Fourier space, gives
\begin{equation}
\widehat{C}_{RR} = 2R^{-2}\widehat{C}+\frac{5}{2}\widehat{C}_2+\frac{1}{4}R^2\widehat{C}_4.
\end{equation}
We can therefore express the variance $\sigma_{RR}^2$ as
\begin{equation}
\sigma_{RR}^2 = 4 R^{-4}\sigma_0^2 + 10 R^{-2}\sigma_1^2 + \frac{35}{2}\sigma_2^2+\frac{5}{4}R^2\sigma_3^2+\frac{1}{16}R^4\sigma_4^2.
\end{equation}
Finally, the correlation factors are given in terms of moments of the power spectrum as,
\begin{equation*}
\gamma_{0,2} = \frac{\langle C C_2 \rangle}{\sigma_0\sigma_2} = - \frac{\sigma_1^2}{\sigma_0\sigma_2},
\end{equation*}
\begin{equation*}
\gamma_{0,RR} = \frac{\langle C C_{RR} \rangle }{\sigma_0\sigma_{RR}} = \frac{2R^{-2}\sigma_0^2-\frac{5}{2}\sigma_1^2+\frac{1}{4}R^2\sigma_2^2}{\sigma_0\sigma_{RR}},
\end{equation*}
\begin{equation}
\gamma_{2,RR} = \frac{\langle C_2 C_{RR} \rangle}{\sigma_2\sigma_{RR}} = \frac{2R^{-2}\sigma_1^2+\frac{5}{2}\sigma_2^2+\frac{1}{4}R^2\sigma_3^2}{\sigma_2\sigma_{RR}}.
\end{equation}

We now turn our attention to the possible range of values for the correlation factors. To achieve this, we will consider a range of shapes for the (smoothed) power spectrum parameterised as
\begin{equation}
\mathcal{P}_{\delta_R} = \mathcal{A} \left( k R \right)^4 \exp\left(-\frac{(kR)^2}{2}\right)\exp\left( -\frac{\mathrm{ln}(k/k_*)^2}{2\Delta^2} \right),
\label{eqn:powerSpectrumPeak}
\end{equation}
where $\mathcal{A}$ describes the amplitude, the $(kR)^4$ term describes the super-horizon growth of density perturbations, and the first exponential term is the smoothing term. The second exponential term describes a lognormal peak, with a peak at $k_*$ and width $\Delta$.  Note that, due to the $k^4$ growth at small $k$ and exponential smoothing at large $k$, the power spectrum will inevitably be relatively narrow, regardless of the choice of $\Delta$. Varying $\Delta$ from $0$ to $\infty$ describes the minimum and maximum allowed widths respectively.

\begin{figure}
 \centering
  \includegraphics[width=0.7\textwidth]{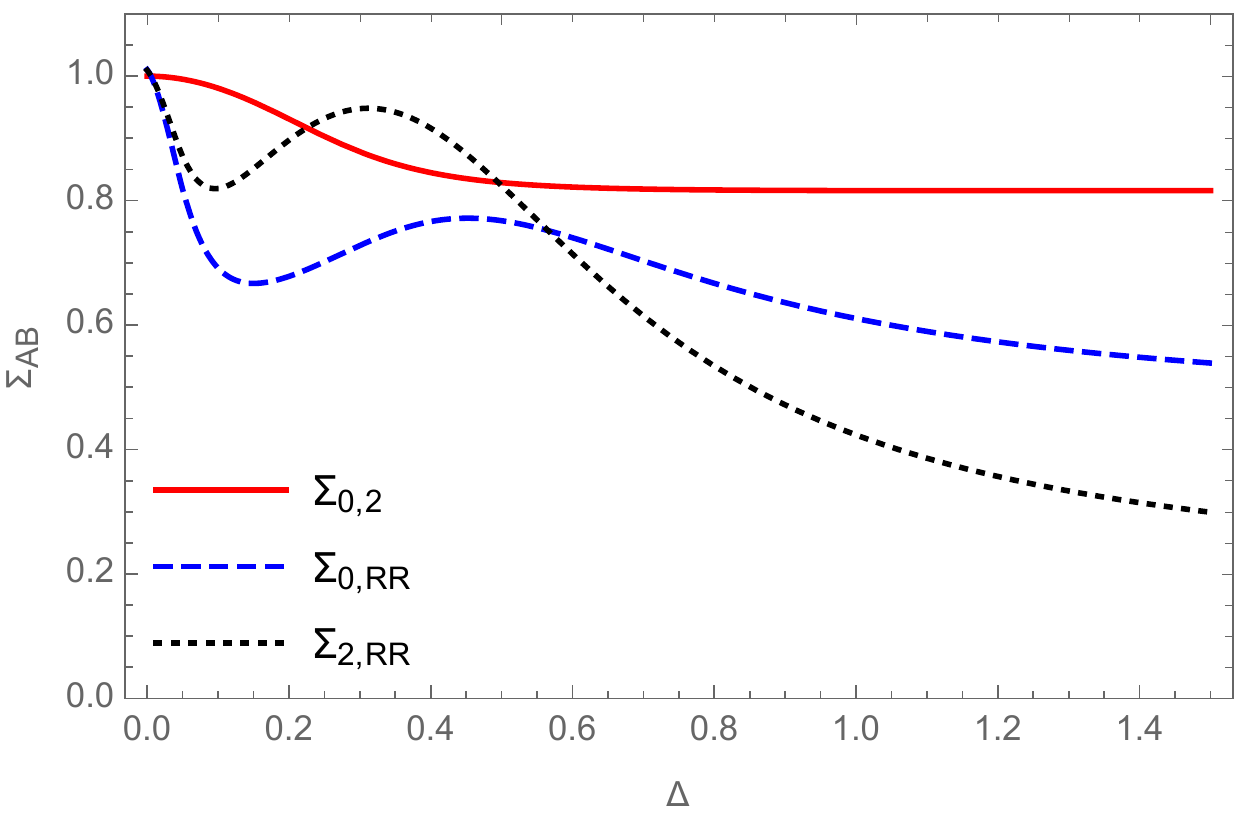} 
  \caption{ The value of the relevant correlation factors are plotted for different widths of the power spectrum $\Delta$, given in equation \eqref{eqn:powerSpectrumPeak}. We have assumed values $R = 1/k_*  = 1$.}
  \label{fig:factors}
\end{figure}

Figure \ref{fig:factors} shows the values for the correlation factors as a function of the power spectrum width $\Delta$. We have used the values $R = 1/k_*  = 1$, although in general using different values does change the value of the correlation factors (though note that the correlation factors are independent of $k_*$ and $R$ as $\Delta \rightarrow 0$ or $\Delta \rightarrow \infty$). As $\Delta\rightarrow0$, all the correlation factors asymptote to unity, and as $\Delta -> \infty$, the correlation factors asymptote to $\gamma_{0,2}=\sqrt{\frac{2}{3}}$, $\gamma_{0,RR}=\frac{\sqrt{2}}{3}$ and $\gamma_{2,RR}=\frac{1}{3\sqrt{3}}$.

\section{The conditional PDF from the 3-variate normal}
\label{app:PDF}

In order to gain a better understanding of the integral over the 3-variate normal PDF seen in Section \ref{sec:npk}, we will briefly discuss the factorization of the conditional PDF.
The 3-variate normal PDF (with zero mean) is given by
\begin{equation}
\mathcal{N}(\mathbf{X}) = \frac{1}{(2\pi)^{3/2}}(\mathrm{det}\Sigma)^{-1/2}\exp\left( -\frac{1}{2}\mathbf{X}^T \Sigma^{-1} \mathbf{X}\right),
\label{eqn:peakDensity}
\end{equation}
where $\mathbf{X} = \{ \nu,J_1,\zeta_{00} \}$ and $\Sigma$ is the covariance matrix, whose elements are given by the cross-correlation coefficients $\gamma_{AB} = \langle C_A C_B \rangle/(\sigma_A\sigma_B)$ defined in equation \eqref{eqn:coeffs}.
For the three variables involved in the PDF, the relevant coefficients are $\gamma_{0,2}$, $\gamma_{0,RR}$ and $\gamma_{2,RR}$.

To diagonalise the PDF we introduce the following normalized Gaussian variables,
\begin{equation}
  x = \frac{J_1 - \gamma_{0,2}\nu}{\sqrt{1-\gamma_{0,2}^2}}
  \quad, \qquad
y = \frac{\zeta_{00} - \gamma_{0,RR}\nu}{\sqrt{1-\gamma_{0,RR}^2}}\,,
\end{equation}
which depend linearly on $J_1$ and $\zeta_{00}$ respectively, but are independent of $\nu$ (since the components that correlate with $\nu$ have been subtracted). Similarly, we define the variable $z$,
\begin{equation}
z = \frac{y - \gamma_{x,y}x}{\sqrt{1-\gamma_{x,y}^2}}\,,
\end{equation}
where the correlation factor $\gamma_{x,y}$ can be expressed as
\begin{equation}
\gamma_{x,y} = \frac{\gamma_{2,RR}-\gamma_{0,2}\gamma_{0,RR}}{\sqrt{( 1-\gamma_{0,2}^2 ) ( 1-\gamma_{0,RR}^2 )}}\,.
\end{equation}
The variable $z$ depends linearly on $\zeta_{00}$ and  is independent of both $\nu$ and $x$. 
The 3 variables $\nu$, $x$ and $z$ are mutually independent and their PDF automatically factorizes. Therefore, the PDF of $\nu$, $J_1$ and $\zeta_{00}$ can be expressed as
\begin{equation}
  \mathcal{N}(\nu,J_1,\zeta_{00})
  = \frac{\mathcal{N}(\nu)\mathcal{N}(x)\mathcal{N}(z)}{\sqrt{(1-\gamma_{0,2}^2)( 1-\gamma_{0,RR}^2)(1-\gamma_{x,y}^2)}}\,.
\label{eqn:joint}
\end{equation}

Let us now consider the integral over the PDF which appears in equation \eqref{eqn:barN}. Using equation \eqref{eqn:joint}, then substituting in $\bar{\nu}$, $\bar{J}_1 = (4\sigma_0/R^2\sigma_2)\bar{\nu}$ and $\zeta_{00}$ into the expressions for  $\nu$, $x$ and $z$ gives, after some simple algebra,
\begin{multline}
\int\limits_{0}^{\infty}\mathrm{d}\zeta_{00}\left| \zeta_{00} \right|\mathcal{N}(\bar{\nu},\bar{J}_1,\zeta_{00}) = \frac{1}{2\pi\sqrt{\big( 1-\gamma_{0,2}^2 \big)}} \exp\left( - \frac{\left( 1+\frac{16\sigma_0^2}{R^4\sigma_2^2}-\frac{8\sigma_0\gamma_{0,2}}{R^2\sigma_2} \right)\bar{\nu}^2}{2\big( 1-\gamma_{0,2}^2 \big)} \right)\\
\times  \int\limits_{0}^{\infty}\mathrm{d}\zeta_{00}\left| \zeta_{00} \right|  \frac{1}{\sqrt{2\pi\big( 1-\gamma_{0,RR}^2 \big)\big( 1-\gamma_{x,y}^2 \big)}}\exp\left(-\frac{\left(\zeta_{00} -\alpha \bar{\nu} \right)^2}{2\big( 1-\gamma_{0,RR}^2 \big)\big( 1-\gamma_{x,y}^2 \big)} \right),
\label{eqn:PDFintegral}
\end{multline}
where
\begin{equation}
\alpha \equiv \gamma_{0,RR}+ \frac{\gamma_{2,RR}-\gamma_{0,RR}\gamma_{0,2}}{1-\gamma_{0,2}^2}\left( \frac{4\sigma_0}{R^2\sigma_2}-\gamma_{0,2} \right).
\label{eqn:alpha}
\end{equation}
This is a relatively complicated formula, but in the next section we will discuss the high-peak limit relevant for PBH formation, where the benefit of expressing the PDF in this form will become clear.

\section{The high-peak approximation}
\label{sec:highPeak}

The expression for the average peak number-density is given by equation \eqref{eqn:barN}, with the function $f(J_1)$ defined in equation \eqref{eqn:f} and an expression for the integral given in equation \eqref{eqn:PDFintegral}. However, if $\nu$ is large, as is generally the case when considering PBH formation, this can be significantly simplified.

Firstly, the function $f(J_1)$ quickly asymptotes to $J_1^3$ as $J_1$ becomes large, and so we have
\begin{equation}
f(\bar{J}_1) \approx \frac{64\sigma_0^3}{R^6\sigma_2^3}\bar{\nu}^3.
\end{equation}

Secondly, we will consider the integral over $v_{RR}$ in \eqref{eqn:PDFintegral}. In the limit $\bar{\nu}\rightarrow \infty$, we can take the integral to be an integral over a Dirac-delta function, centred on $\alpha\bar{\nu}$, multiplied by $\left| v_{RR} \right|$.
\begin{multline}
 \int\limits_{0}^{\infty}\mathrm{d}\zeta_{00}\left| \zeta_{00} \right|  \frac{1}{\sqrt{2\pi\left( 1-\gamma_{0,RR}^2 \right)\left( 1-\gamma_{2,RR}^2 \right)}}\exp\left(-\frac{\left(\zeta_{00} -\alpha \bar{\nu} \right)^2}{2\left( 1-\gamma_{0,RR}^2 \right)\left( 1-\gamma_{2,RR}^2 \right)} \right) \\
 \approx  \int\limits_{0}^{\infty}\mathrm{d}\zeta_{00}\left| \zeta_{00} \right| \delta_{\mathrm{D}}\left( \zeta_{00}-\alpha \bar{\nu} \right) = \alpha \bar{\nu}.
\end{multline}

Substituting this into equation \eqref{eqn:barN}, we arrive at the following expression for the number density of high peaks,
\begin{equation}
n_{\mathrm{hi-pk}} (\nu)= \frac{16\sqrt{2}}{3^{3/2}\pi^{5/2}}\frac{\sigma_{RR}\sigma_0^3}{\sigma_2\sigma_1^3\sqrt{1-\gamma_{0,2}^2}}\alpha \nu^4\exp\left( -\frac{1}{2} \frac{\left( 1+\frac{16\sigma_0^2}{R^4\sigma_2^2}-\frac{8\sigma_0\gamma_{0,2}}{R^2\sigma_2}  \right)\nu^2}{1-\gamma_{0,2}^2} \right),
\end{equation}
where the factor $\alpha$ is given in equation \eqref{eqn:alpha}, and is typically of order unity, except for very narrow power spectra, where it becomes large.

\section{The abundance and mass function of primordial black holes}
\label{app:abundance}

While we will not consider further the details of the calculation, the formulae needed to calculate the PBH abundance and mass function are included here for completeness. The mass fraction of the universe collapsing to form PBHs at a specific time can be calculated by integrating over $n_{\mathrm{hi-pk}}$, given in equation \eqref{eqn:hiPeak},
\begin{equation}
\beta(M_{\mathrm H}) = \int\limits_{\delta_c}^{\infty}\mathrm{d}C \mathcal{K}\left( C - C_c \right)^\gamma n_{\mathrm{hi-pk}}\left( \frac{C}{\sigma_0(M_\mathrm{H})} \right),
\end{equation}
where the $\mathcal{K}\left( C - C_c \right)^\gamma$ accounts for the different mass PBHs forming from different amplitude perturbations and we have used $\nu = C/\sigma_0$.
Secondly, we can calculate the total abundance of PBHs, in terms of the PBH density parameter at matter-radiation equality, by integrating over all of the times at which PBH formation occurs (in this case parameterized by the horizon mass),
\begin{equation}
\Omega_{\mathrm PBH} = \int\limits_{M_{min}}^{M_{max}}\mathrm{d}\left( \mathrm{ln}M_\mathrm{H}\right) \left( \frac{M_{\mathrm eq}}{M_{\mathrm H}} \right)^{1/2}\beta(M_{\mathrm H}),
\label{eqn:densityParameter}
\end{equation}
where the $M_{\mathrm eq}/M_{\mathrm H}$ accounts for the red-shift of matter density relative to radiation density during radiation domination (where we have assumed complete radiation domination until the time of equality). Finally, there are several ways of defining the mass function, involving a derivative with respect to the PBH mass,
\begin{equation}
f\left(M_{\mathrm PBH}\right) = \frac{1}{\Omega_{\mathrm CDM}}\frac{\mathrm{d}\Omega_{\mathrm PBH}}{\mathrm{d}(\mathrm{ln}M_{\mathrm PBH})},
\end{equation}
which is normalised such that $\int\mathrm{d}(\mathrm{ln}M_{\mathrm PBH})f\left(M_{\mathrm PBH}\right) = \Omega_{\mathrm PBH}/\Omega_{\mathrm CDM}$, and 
\begin{equation}
\psi\left(M_{\mathrm PBH}\right) =  \frac{1}{\Omega_{\mathrm PBH}}\frac{\mathrm{d}\Omega_{\mathrm PBH}}{\mathrm{d}M_{\mathrm PBH}},
\end{equation}
which is normalised such that $\int\mathrm{d}M_{\mathrm PBH}\psi\left(M_{\mathrm PBH}\right) = 1$. In the case of a broad (scale-invariant) power spectrum, $\beta$ is constant over $M_\mathrm{H}$, in which case $\psi\left(M_{\mathrm PBH}\right)\propto M_{\mathrm PBH}^{-3/2}$ (as found in the recent paper \cite{DeLuca:2020ioi}), with a factor $M_{\mathrm PBH}^{-1/2}$ coming from the red-shift term in equation \eqref{eqn:densityParameter}, and a further factor $M_{\mathrm PBH}^{-1}$ coming from the derivative with respect to $M_{\mathrm PBH}$.



\bibliographystyle{JHEP} 
\bibliography{bibfile}

\end{document}